\documentclass[aps,preprint]{revtex4}%
\usepackage{amsfonts}
\usepackage{amsmath}
\usepackage{amssymb}
\usepackage{graphicx}%
\providecommand{\U}[1]{\protect\rule{.1in}{.1in}}

\begin{document}
\preprint{ }
\title{{\huge Primordial Nucleosynthesis and Finite Temperature QED}}
\author{Mahnaz Q. Haseeb$^{\ast}$, Obaidullah Jan, and Omair Sarfaraz}
\affiliation{\textit{Department of Physics, COMSATS Institute of Information Technology,
Park Road, Chak Shahzad, Islamabad, Pakistan }$^{\ast}%
{\small mahnazhaseeb@comsats.edu.pk}$}
\keywords{QED at finite temperature, corrections to He abundance parameter}
\pacs{PACS: 11.10.Wx, 26.35.+c, 12.20.-m, 14.60.Cd}

\begin{abstract}
Abundances of light nuclei formed during primordial nucleosynthesis are
predicted by Standard Big Bang Model. The latest data from WMAP, with
precision's higher than ever before, provides a motivation to determine
theoretical higher order corrections to change in helium abundance and the
related parameters. Here we evaluate the QED corrections to the change in
these parameters during primordial nucleosynthesis using finite temperature
effects at the two loop level. Relative variations in neutron decay rate,
total energy density of the universe, relative change in neutrino temperature
etc., with two loop corrections to electron mass, at the timescale when QED
corrections were relevant, have been estimated.

\end{abstract}
\maketitle
\homepage{ }
\affiliation{}

\section{\qquad Introduction}

Big Bang cosmology \cite{Weinberg2004}, combined with the observational data
from recent satellites and experimental data analysis available from
large-scale particle colliders, results in a confident extrapolation of the
history of universe with unprecedented precision. Cosmic Background Explorer
(COBE) and Wilkinson Microwave Anisotropy Probe (WMAP) \cite{Hinshaw2009} have
been providing observational input for modifications to light element
abundances, at the time of Big Bang Nucleosynthesis (BBN)
\cite{AlphBethGam1948}, through their sophisticated observational techniques.
WMAP was specifically launched to confirm and reinforce the understanding of
standard model of cosmology and precisely determine the cosmological
parameters. It was stunningly successful during its seven years of operation
providing, precision up to five decimal places \cite{Bennet2011}. Latest
observational missions, Planck and Herschel are further providing data with
even more sensitivity, opening newer windows for understanding, how the
universe evolved. There has been a persistent effort to re-examine and refine
the possibilities of improved theoretical input to these parameters.

BBN provides one of the earliest direct cosmological probes on parameters of
the universe when it was at a temperature scale of around MeV ($\thicksim
10^{10}K$). The role of BBN in formation of primordial light elements has been
extensively discussed in literature \cite{Cyburt2003}-\cite{Bailin2004}.
Helium-4 ($^{4}He$) yield is sensitive to the expansion rate of the early
universe. Very early universe contained highly energetic relativistic
particles, i.e., photons, electrons, positrons and neutrinos. At such high
energies the weak interactions given below played a major role
\cite{Hecht1971}-\cite{Sears1964} in regulating the relative number of protons
and neutrons:%

\[
n\rightleftharpoons p+e^{-}+\overline{\nu}_{e},
\]

\[
p+e^{-}\rightleftharpoons n+\nu_{e},
\]

\[
n+e^{+}\rightleftharpoons p+\overline{\nu}_{e},
\]

\[
\text{ }n+p\rightleftharpoons D+\gamma.
\]

During the primordial era of nucleosynthesis, temperature was high enough for
the mean energy per particle to be greater than the binding energy of $D$
($2.2$ MeV) thereby any $D$ formed was immediately destroyed. Thus formation
of $^{4}He$ was delayed until the universe became cool enough to form $D$ (at
$T\sim0.1$ MeV), when there was a sudden burst of nuclei of light elements
production. The abundances of light elements got fixed and only changed as
some of the radioactive products of BBN such as $^{3}He$ decayed. Theory of
BBN provides a detailed description on production of $D,$ $^{3}He,$ $^{4}He$,
and $^{7}Li$ as well as precise quantitative predictions for their abundances.
The theory predicts mass abundances of about 75\% of $^{1}H$, about 25\%
$^{4}He$, about 0.01\% of $D$, traces ($\thicksim$ $10^{-10}$) of $Li$ and
$Be$, and no other heavy elements. Then until the first star formation, there
were no significantly produced new elements.

Changes brought about by cosmic expansion are combined with thermodynamics to
calculate the fraction of protons and neutrons based on temperature. $^{4}He$
abundance is important because $^{4}He$ in the universe is much more than what
could be explained by stellar nucleosynthesis. For $^{4}He$, there is a good
agreement of predictions from BBN with observations. Standard Big Bang
Nucleosynthesis (SBBN) gives a measure for $^{4}He$ abundance $Y_{p}=0.25$
which is taken equivalent to $^{4}He$ abundance on the basis that all the
neutrons wind up into $^{4}He$ because of its stability. Primordial $^{4}He$
abundance $Y=0.241\pm\ 0.006$ leads to a very small baryon density parameter
$\eta_{_{10}}<3$ \cite{Miele2009} (with $\eta_{_{10}}=10^{10}\eta_{b}$ where
$\eta_{b}=\frac{N_{b}}{N_{\gamma}}$) which is largely inconsistent with what
was predicted by SBBN. $^{4}He$ abundance deduced from WMAP results, as a
function of the baryonic density with Monte-Carlo calculation, gives $Y=$
$0.2476\pm0.0004$ \cite{Coc2010}. The data from WMAP showed that if the Big
Bang creation model is correct, the value based on Cosmic Microwave Background
(CMB) prediction\ is $Y=0.24819_{-0.00040}^{+0.00029}\pm0.0006$ (syst.)
\cite{Spergel2007}.

The perturbation in $^{4}He$ abundance during the era of primordial
nucleosynthesis is related to mass shift arising from radiative corrections to
particles propagating in the early universe \cite{Hecht1971}-\cite{Sears1964}.
Abundances of light elements formed in the early universe are influenced by
finite temperature effects. The temperature range particularly of relevance in
primordial nucleosynthesis of light elements is the range where finite
temperature Quantum Electrodynamics (QED) interactions are valid. At these
temperatures virtual electron-positron pairs couple to photons in loops, and
vice versa, affecting particle dispersion. One loop corrections to QED
processes have been calculated in detail at finite temperature and by
including densities, in some cases, for various physical environments
\cite{Levinson1985}-\cite{Masood1992b}. During primordial nucleosynthesis, the
modification due to temperature prevailed over density effects with
$\mu/T\leqslant10^{-9}$, where $\mu$\ is the chemical potential of the
particles such as electrons, neutrinos and their anti particles present in the background.

We used real time formalism because the temperature corrections with this
formalism are obtained as separate terms additive to the zero temperature
part. At finite temperature, the corrections are usually calculated for
limiting cases of temperature $T\ll m$ and $T\gg m,$ where $m$\ is the
electron mass. These were re-examined in a general form so that the range of
threshold temperatures ($T$ $\thicksim$ $m$) for the creation of electron
positron pairs are also included \cite{Ahmed1987a}. The ranges of temperature
$T<m$ and $T>m$ can be retrieved as limiting cases. The relative change in
electron mass and the variation in Helium abundance parameter can be
calculated vs $T/m$. Higher order in $\alpha$\ modifications to the electron
self mass are worth estimating for even finer corrections.

\section{\qquad Self Energy of Electron at Finite Temperature}

Quantum Field Theory assumes that particles are analogous to excitations of a
harmonically oscillating field permeating space-time. The state of vacuum is
the absence of particles, but it is not devoid of energy and fields. Finite
temperature effects are considered in heat bath containing hot particles and
antiparticles which can mediate interactions between real and virtual
particles. Particle propagation in vacuum can be taken as the propagation with
such interactions switched off. Therefore, properties of system with
background medium are somewhat different from the system in which all the
particles propagate freely. In finite temperature environments, relevant in
QED, electrons and photons propagate in statistical background at energies
around thresholds for the production\ of electron-positron pairs. The
temperature effects that arise due to continuous particle exchanges during the
physical interactions in a heat bath need to be appropriately taken into consideration.

Calculations involved in finite-temperature field theory are similar to those
in perturbative quantum field theory at $T=0$. The interactions taking place
when photons propagate through a system in the presence of fermions in the
background, or vice versa, contribute to perturbative corrections in
dynamics\ describing the system. The statistical effects of particles
propagating in a heat bath of photons, electrons and positrons at finite
temperature enter the theory through Fermi-Dirac and Bose-Einstein
distribution functions. These interactions with the background heat bath are
incorporated by modifying the particle propagators. Using the finite
temperature formulations, scattering amplitudes and loop corrections are
calculated. From poles of the propagators, modified dispersion relations are
obtained \cite{Donoghue1983}.

Self energies of particles acquire temperature corrections in a heat bath due
to the possibility of energy-momentum exchanges with real particles. Thermal
mass is radiatively generated in such particle interactions and serves as a
kinematical cut-off, in production rate of particles from the heat bath.
Electrons and photons acquire dynamically generated mass due to plasma
screening through the self energy corrections at finite temperature. This
leads to modification in masses, coupling etc., of interacting particles at
finite temperature. The electromagnetic properties of the medium in which they
propagate are also influenced in such background. Physical mass of electron at
one loop level was obtained from electron self-energy \cite{Donoghue1983} for
$T\ll m$ and $T\gg m$. Using the corrections to electron self energy, first
order in $\alpha\ $corrections to primordial parameters have been determined
\cite{Dicus1983} for these limits of temperature. In particular from the point
of view of light elements abundances at the time of primordial
nucleosynthesis, self energy corrections to electron propagators have been of
significance even at $T$ $\thicksim$ $m$ \cite{Saleem1987}. Corrections to
self energies in QED have been calculated at the two loop level
\cite{Qader1991}-\cite{Haseeb2011}, in real time formalism.

\section{\qquad Modified Electron Mass at Finite Temperature}

During primordial nucleosynthesis the relative changes in Helium abundance
parameter, neutron decay rate, energy density, etc. have been shown to depend
on relative shift in electron mass \cite{Hecht1971}-\cite{Sears1964}.
QED\ finite temperature corrections obtained by Dicus \textit{et al.}
\cite{Dicus1982} for one loop were already included in the BBN codes
\cite{Walker1991}. The relative shift in electron mass was calculated in
detail \cite{Qader1992} at order $\alpha^{2}$\ by one particle reducible and
one particle irreducible self energy diagrams with two loops. This shift in
electron mass up to second order in $\alpha$\ was calculated in a general form
in which $T\thicksim m$ was also inclusively represented. The results for
$T\ll m$ and $T>m$ were retrievable as limiting cases. The mass shift at
finite temperature was calculated, which leads to the determination of the
physical mass of an electron: $m_{phy}=m+\delta m^{(1)}+\delta m^{(2)}$. Here
$m\ $is the electron mass without temperature background and $\delta m^{(1)}$
and $\delta m^{(2)}$\ are the shifts in electron mass due to temperature
effects at one and two loop level respectively. One particle reducible
corrections with two loop were presented \cite{Qader1992} to obtain an
expression for $\frac{\delta m}{m}$. The expression for self energy for one
particle reducible diagrams was taken as:
\begin{equation}
\left(  \frac{\delta m}{m}\right)  ^{n}\simeq\left(  \frac{\alpha\pi T^{2}%
}{2m^{2}}\right)  ^{n},\label{nLoops}%
\end{equation}
where $n$ represents the number of loops under consideration. Thus, on
iteration of one loop result, the relative change in electron mass for high
temperature \cite{Ahmed1987a} is
\begin{equation}
\left(  \frac{\delta m}{m}\right)  ^{2}\simeq\frac{\alpha^{2}\pi^{2}T^{4}%
}{4m^{4}}.\label{2loopT>m}%
\end{equation}
In case of one particle irreducible diagram for two loops, the leading term
for temperature values around the electron mass was calculated
\cite{Qader1992} to be:
\begin{equation}
\frac{\delta m^{(2)}}{m}\overset{T\sim m}{\longrightarrow}-\frac{16\alpha^{2}%
}{\pi^{2}}\varsigma(3)\left(  \frac{T}{m}\right)  ^{3}\left(  \frac{E}%
{m}\right)  ^{2}.\label{2loopTorderM}%
\end{equation}
The expression for the two loop shift in electron mass at finite temperature
$\frac{\delta m^{(2)}}{m}$\ was derived \cite{Haseeb2011}. From ref.
\cite{Masood2012}, at two loop level,\ the leading order contributions to the
electron mass at low temperature $(T<m)$, contributions to $\frac{\delta m}%
{m}$ come out to be:%

\begin{equation}
\frac{\delta m^{(2)}}{m}\overset{T<m}{\longrightarrow}15\alpha^{2}\left(
\frac{T^{2}}{m^{2}}\right)  . \label{2loopIrredT<m}%
\end{equation}
On the other hand, similar contributions at high temperature $(T>m)$
\cite{Masood2012} are given by:%

\begin{equation}
\frac{\delta m^{(2)}}{m}\overset{T>m}{\longrightarrow}\frac{\alpha^{2}}%
{4}\left[  \pi^{2}\left(  \frac{T^{2}}{m^{2}}\right)  ^{2}-\frac{m^{2}}{T^{2}%
}\right]  . \label{2loopIrredT>m}%
\end{equation}

Using the expression for relative change in electron mass in eqs.
(\ref{2loopT>m})-(\ref{2loopIrredT>m}), in various temperature limits, one can
determine the temperature effect on the primordial $^{4}He$ abundance
parameter $\Delta Y$ at the two loop level.

\section{\qquad Results and Discussion}

The presence of finite temperature background at the time of synthesis of
light nuclei makes it relevant to include temperature corrections to electron
mass while determining the variation in primordial $^{4}He$ abundance. We have
determined here the effect of two loop correctionsin QED to the change in
Helium abundance parameter $\Delta Y$, related to the relative shift in
electron mass \cite{Schwarchild1958}-\cite{Sears1964}, \cite{Levinson1985}%
-\cite{Donoghue1983} $\frac{\delta m}{m}$ arising from temperature dependent
radiative corrections. For this, the relevant relations needed are:%

\[
\Delta Y=-0.2\frac{\Delta\lambda}{\lambda},
\]
such that $\frac{\Delta\lambda}{\lambda}$ is relative change in neutron decay
rate given by:%

\[
\frac{\Delta\lambda}{\lambda}=-0.2\left(  \frac{m}{T}\right)  ^{2}\frac{\delta
m}{m},
\]
where $\lambda=\left(  \tau_{1/2}\right)  ^{-1}$ with $m$ as the mass of the
propagating particle and $T$ the temperature of background heat bath.
Therefore $\Delta Y$ becomes
\begin{equation}
\Delta Y=0.04\left(  \frac{m}{T}\right)  ^{2}\frac{\delta m}{m}.
\label{deltaY}%
\end{equation}

The interesting range of QED temperatures for primordial $^{4}He$%
\ synthesis\ is from $T\thicksim0.1$ MeV$\ $to a few MeV. One loop corrections
estimated \cite{Saleem1987} were $\Delta Y=0.4\times$ $10^{-3}$ at $T$
$\thicksim$ $m$ which falls to $0.3\times$ $10^{-3}$ at $T$ $\thicksim$ $m/3$.
A rough estimate was earlier given \cite{Haseeb2011}, using eq.
(\ref{2loopTorderM}), at energies $\thicksim1$ MeV from one particle
irreducible diagram. We now further analyze the dependence of $\frac{\delta
m^{(2)}}{m}$\ on finite temperature corrections for the temperature ranges of
interest during the primordial\ synthesis of elements.%

\begin{figure}[ptb]%
\centering
\includegraphics[
height=1.6405in,
width=2.6083in
]%
{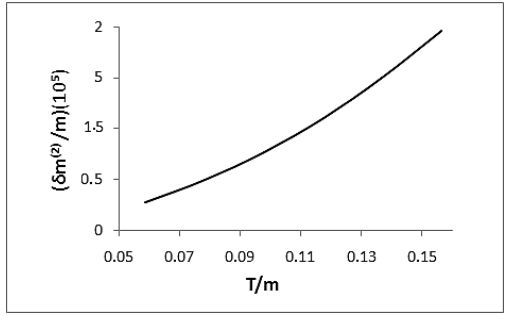}%
\caption{$\frac{\delta m^{(2)}}{m}$\ vs $T/m$\ for $T<m,$ for two loop one
particle irreducible diagrams. }%
\label{TltMDelMbyM}%
\end{figure}

The temperatures of particular interest during BBN are $0.08$ MeV $\lesssim
T\lesssim0.03$ MeV, since most of the Helium produced, was synthesized during
this range of temperature. Two loop correction at temperature $T<m,$ from one
particle irreducible diagrams, the value of $\Delta Y$ is $3.20\times10^{-5}$.
A similar situation was found while determining these ratios for one loop
corrections in ref. \cite{Saleem1987}. This value remains constant here since
the $T/m$\ dependence cancels out in the two loop correction to expression for
$\Delta Y,$ for low temperatures, as can be seen from eqs.
(\ref{2loopIrredT<m}) and (\ref{deltaY}). It is evident that the two loop
corrections at temperature $T<m$ though small, are not completely negligible.
The corrections to $\Delta Y$ obtained here are in the fifth decimal places.
The behavior of $\frac{\delta m^{(2)}}{m},$ with change in $T/m$ for the
temperature range $0.08$ MeV $\lesssim T\lesssim0.03$ MeV is plotted in figure
\ref{TltMDelMbyM}. The variation in $\frac{\delta m^{(2)}}{m},$ with change in
$T/m$ for this temperature range comes out to be between $2.75\times10^{-6}$
and $1.96\times10^{-5}.$%

\begin{figure}[ptb]%
\centering
\includegraphics[
height=1.606in,
width=2.623in
]%
{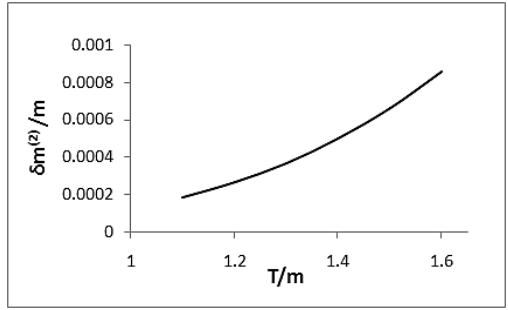}%
\caption{$\frac{\delta m^{(2)}}{m}$\ vs $T/m$\ for $T>m,$ for two loop one
particle irreducible diagrams. }%
\label{TgrMDelMbyM}%
\end{figure}

Using eq. (\ref{2loopT>m}), the variation in $\frac{\delta m^{(2)}}{m},$ with
change in $T/m$ for the temperature range relevant in QED soon after the
freezeout of weak interactions, is between $1.8\times10^{-4}$ and
$8.6\times10^{-4}.$ This variation in $\frac{\delta m^{(2)}}{m}$ vs $T/m$\ for
$T>m$ in this temperature range is plotted in figure \ref{TgrMDelMbyM}, for
two loop one particle irreducible diagrams. 

\bigskip%
\begin{figure}[ptb]%
\centering
\includegraphics[
height=1.7902in,
width=2.9732in
]%
{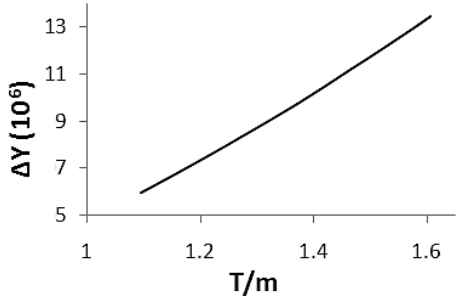}%
\caption{$\Delta Y$\ vs $T/m$\ for $T>m,$ for two loop one particle
irreducible diagrams for $0.82$ MeV $\lesssim T\lesssim0.56$ MeV. }%
\label{DelY}%
\end{figure}

The plot of $\Delta Y$ vs $T/m$ for temperatures $0.82$ MeV $\lesssim
T\lesssim0.56$ MeV, is given in figure \ref{DelY}. The values of $\Delta Y$
vary between\ $1.35\times10^{-5}$ and $5.95\times10^{-6}$. These values are
somewhat smaller at $T>m$ probably due to the reason that even the synthesis
of light elements was not much as compared to this synthesis during the range
of temperature $0.08$ MeV $\lesssim T\lesssim0.03$ MeV, when there was sudden
burst of light elements nucleosynthesis. The values of $\Delta Y$ obtained for
two loops are significantly smaller than $\Delta Y$ values with one loop
corrections. This is in accordance with the perturbative nature of QED. It
needs to be mentioned that from eq. (\ref{2loopTorderM}), the value of $\Delta
Y\thicksim4.15\times$ $10^{-6}$ at $T$ $\thicksim$ $m$, (energies
$\thicksim0.5\ $MeV). 

Other parameters such as the relative change in the neutrino temperature
\[
\frac{\Delta T_{\nu}}{T_{\nu}}=-0.1\left(  \frac{m}{T}\right)  ^{2}%
\frac{\delta m}{m},
\]
the change in neutrino temperature relative to photon temperature%

\[
\frac{\Delta T_{\nu}}{T}=-\left(  \frac{m}{5T}\right)  ^{2}\frac{\delta m}%
{m},
\]
and the relative variation in the total energy density of the universe during
the era under consideration%

\[
\frac{\Delta\rho_{T}}{\rho_{T}}=-\left(  \frac{m}{4T}\right)  ^{2}\frac{\delta
m}{m}%
\]
are also calculated using \cite{Dicus1982} the expressions of two loop
modifications to $\frac{\delta m}{m}$ for the relevant temperature limits in
eqs. (\ref{2loopT>m}) - (\ref{2loopIrredT>m}) for two loop irreducible
diagrams. The values from these relations for $T>m$\ are listed in Table 1.

\begin{tabular}
[b]{||c|c|c|c|c|c|c||}\hline\hline
$T$ (MeV) & $\frac{\delta m}{m}$ & $\Delta Y$ & $\frac{\Delta T_{\nu}}{T_{\nu
}}$ & $\frac{\Delta T_{\nu}}{T}$ & $\frac{\Delta\lambda}{\lambda}$ &
$\frac{\Delta\rho_{T}}{\rho_{T}}$\\\hline
$0.6$ & $0.00024$ & $6.96\times10^{-6}$ & $-1.74\times10^{-5}$ &
$6.96\times10^{-6}$ & $-3.5\times10^{-5}$ & $1.09\times10^{-5}$\\\hline
$0.7$ & $0.00046$ & $9.71\times10^{-6}$ & $-2.43\times10^{-5}$ &
$9.71\times10^{-6}$ & $-4.9\times10^{-5}$ & $1.52\times10^{-5}$\\\hline
$0.8$ & $0.00078$ & $1.28\times10^{-5}$ & $-3.19\times10^{-5}$ &
$1.28\times10^{-5}$ & $-6.4\times10^{-5}$ & $2.0\times10^{-5}$\\\hline
$0.9$ & $0.0013$ & $1.62\times10^{-5}$ & $-4.06\times10^{-5}$ & $1.62\times
10^{-5}$ & $-8.1\times10^{-5}$ & $2.54\times10^{-5}$\\\hline
$1.0$ & $0.0019$ & $2.01\times10^{-5}$ & $-5.02\times10^{-5}$ & $2.01\times
10^{-5}$ & $-0.0001$ & $3.14\times10^{-5}$\\\hline
$1.1$ & $0.0028$ & $2.43\times10^{-5}$ & $-6.08\times10^{-5}$ & $2.43\times
10^{-5}$ & $-0.00012$ & $3.80\times10^{-5}$\\\hline
$1.2$ & $0.0039$ & $2.9\times10^{-5}$ & $-7.24\times10^{-5}$ & $2.9\times
10^{-5}$ & $-0.00014$ & $4.52\times10^{-5}$\\\hline
$1.3$ & $0.0055$ & $3.4\times10^{-5}$ & $-8.49\times10^{-5}$ & $3.4\times
10^{-5}$ & $-0.00017$ & $5.31\times10^{-5}$\\\hline
$1.4$ & $0.0074$ & $3.9\times10^{-5}$ & $-9.85\times10^{-5}$ & $3.9\times
10^{-5}$ & $-0.0002$ & $6.16\times10^{-5}$\\\hline
$1.5$ & $0.0097$ & $4.5\times10^{-5}$ & $-0.00011$ & $4.5\times10^{-5}$ &
$-0.00023$ & $7.07\times10^{-5}$\\\hline\hline
\end{tabular}

Table 1. The values of $\Delta Y$, $\frac{\Delta T_{\nu}}{T}$, $\frac
{\Delta\lambda}{\lambda}$, $\frac{\Delta\rho_{T}}{\rho_{T}}$,\ and
$\frac{\Delta T_{\nu}}{T_{\nu}}$ obtained for temperature range $0.6$ MeV
$\leq T\leq$ $1.5$ MeV. 

\bigskip%

\begin{figure}[ptb]%
\centering
\includegraphics[
height=2.9507in,
width=4.9969in
]%
{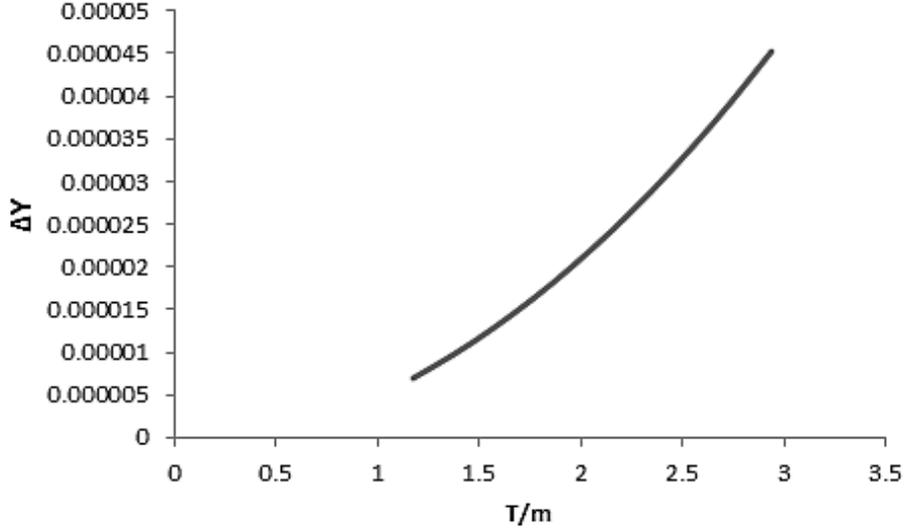}%
\caption{Plot of second order corrections to $\Delta Y$ vs $T/m$ for $0.6$ MeV
$\leq T\leq$ $1.5$ MeV. }%
\end{figure}

The values of $\Delta Y$, $\frac{\Delta T_{\nu}}{T}$, $\frac{\Delta\lambda
}{\lambda}$, $\frac{\Delta\rho_{T}}{\rho_{T}}$,\ and $\frac{\Delta T_{\nu}%
}{T_{\nu}}$ obtained for $T>m$\ in the temperature range $0.6$ MeV $\leq
T\leq$ $1.5$ MeV are plotted vs $T/m$ in figures (4) - (8) respectively. In
figure 4, the second order correction to $\Delta Y$ parameter is plotted vs
$T/m$ . The values of $\Delta Y$ vary from $6.96\times10^{-6}$ to
$4.53\times10^{-5}$ for the range of temperature $0.6$ MeV $\leq T\leq$ $1.5$
MeV. %

\begin{figure}[ptb]%
\centering
\includegraphics[
height=2.8677in,
width=4.9562in
]%
{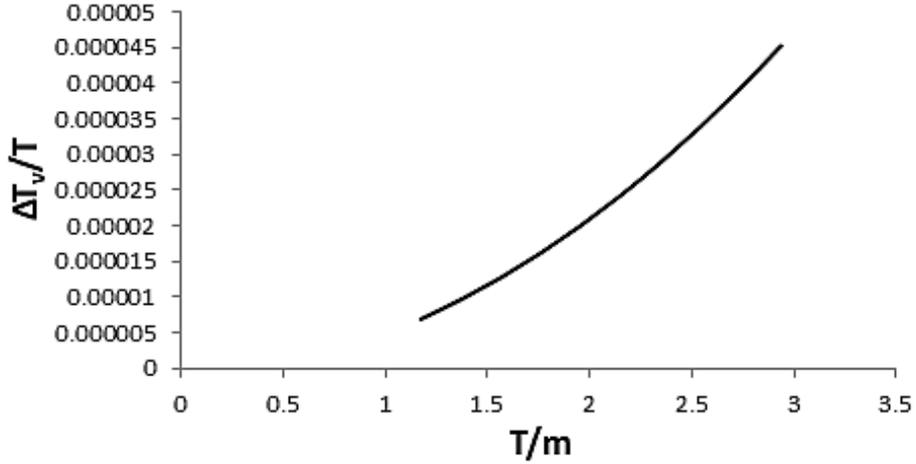}%
\caption{Plot of two loop corrections to $\frac{\Delta T_{\nu}}{T}$ vs $T/m$
for $T>m$.}%
\end{figure}

As portrayed in figure 5 for $\frac{\Delta T_{\nu}}{T},$ the values show an
increasing behavior with increase in $T/m$. Table 1 shows that as $T$ varies
between $0.6$ MeV $\leq T\leq$ $1.5$ MeV, the values of $\frac{\Delta T_{\nu}%
}{T}$ increase from $6.96\times10^{-6}$ to $4.53\times10^{-5}.$

%

\begin{figure}[ptb]%
\centering
\includegraphics[
height=2.7138in,
width=5.0246in
]%
{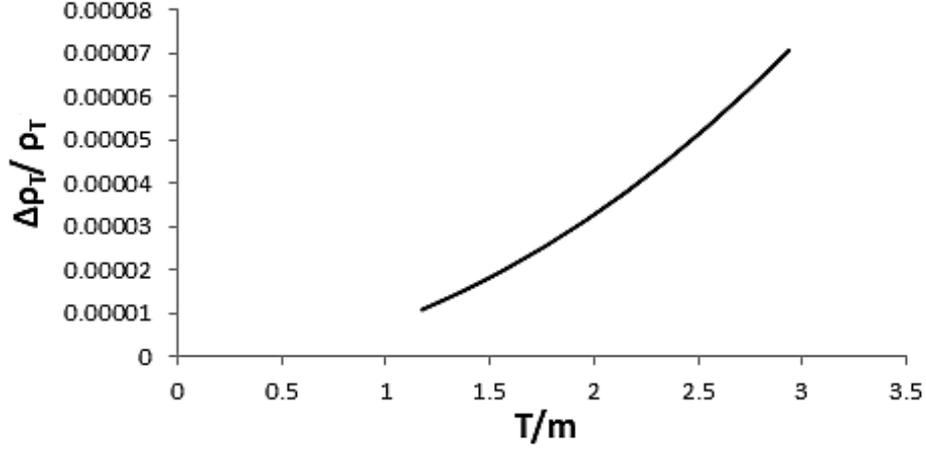}%
\caption{Plot of second order corrections to $\frac{\Delta\rho_{T}}{\rho_{T}}$
vs $T/m$ for $0.6$ MeV $\leq T\leq$ $1.5$ MeV.}%
\end{figure}

Figure 6 also depicts an increase in the values of $\frac{\Delta\rho_{T}}%
{\rho_{T}}$ vs $T/m$. The range in variation of the values of $\frac
{\Delta\rho_{T}}{\rho_{T}}$ is from $1.09\times10^{-5}$ to $7.07\times10^{-5}$
for $0.6$ MeV $\leq T\leq$ $1.5$ MeV, shown in Table 3. Figure 7 gives the
variation of relative decay rate of neutrons $\frac{\Delta\lambda}{\lambda}$
with $T/m$ which is negative and varies from $-3.5\times10^{-5}$ to
$-2.3\times10^{-4}$ for the same variation in temperature.

%

\begin{figure}[ptb]%
\centering
\includegraphics[
height=2.8954in,
width=4.9848in
]%
{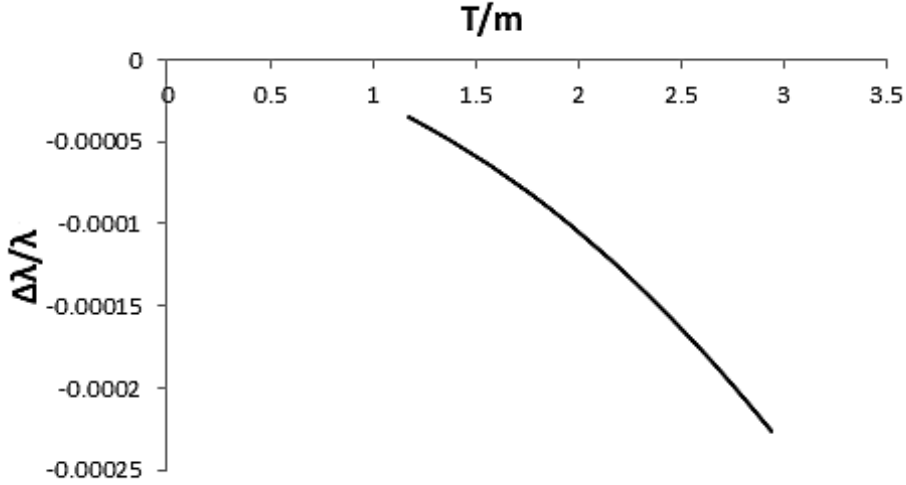}%
\caption{Plot of two loop corrections to $\frac{\Delta\lambda}{\lambda}$ vs
$T/m$ for $T>m$.}%
\end{figure}
%

\begin{figure}[ptb]%
\centering
\includegraphics[
height=2.9507in,
width=4.9415in
]%
{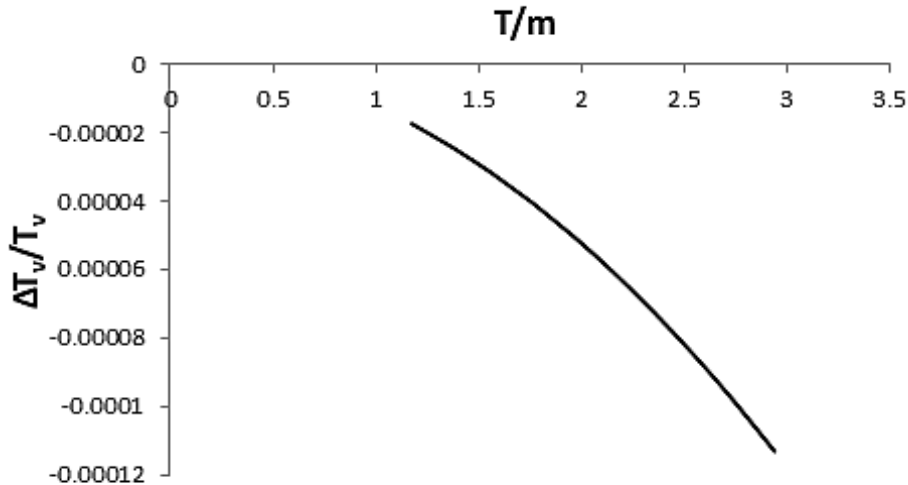}%
\caption{Plot of two loop corrections to $\frac{\Delta T_{\nu}}{T_{\nu}}$ vs
$T/m$ for $T>m$.}%
\end{figure}

Figure 8 is a plot for variation of $\frac{\Delta T_{\nu}}{T_{\nu}}$ from
$-1.74\times10^{-5}$ to $-1.13\times10^{-4}$ for $0.6$ MeV $\leq T\leq$ $1.5$
MeV. The relative variations in all these parameters though small are not
completely negligible.

The latest observational probes such as Planck, Herschel and James Webb Space
Telescope are to provide further fine tuning in precision measurement for the
values of these parameters. The finite temperature corrections around the
electron mass scale are interesting since even up to second order in $\alpha,$
they give corrections within the recently measurable range of observational
probes, within uncertainities. Higher order corrections to weak processes may
be also interesting and useful for estimating modifications to parameters in
the early universe, due to finite temperature effects.

{\small \textbf{Acknowledgement:} The authors are thankful to Prof. Riazuddin
and Samina S. Masood for useful discussions and suggestions. Higher Education
Commission (HEC), Pakistan is acknowledged for partial funding under a
research grant \# 1925 during this work. }

\textbf{Figure Captions}

Figure 1. $\frac{\delta m^{(2)}}{m}$\ vs $T/m$\ for $T<m,$ for two loop one
particle irreducible diagrams.

Figure 2. $\frac{\delta m^{(2)}}{m}$\ vs $T/m$\ for $T>m,$ for two loop one
particle irreducible diagrams.

Figure 3. $\Delta Y$\ vs $T/m$\ for two loop one particle irreducible diagrams
for $0.82$ MeV $\lesssim T\lesssim0.56$ MeV.

Figure 4. Plot of second order corrections to $\Delta Y$ vs $T/m$ for $0.6$
MeV $\leq T\leq$ $1.5$ MeV.

Figure 5. Plot of second order corrections to $\frac{\Delta T_{\nu}}{T}$ vs
$T/m$ for $0.6$ MeV $\leq T\leq$ $1.5$ MeV.

Figure 6. Plot of two loop corrections to$\ \frac{\Delta\rho_{T}}{\rho_{T}}$
vs $T/m$ for $T>m$.

Figure 7. Plot of second order corrections to$\ \frac{\Delta\lambda}{\lambda}$
vs $T/m$ for $T>m$.

Figure 8. Plot of two loop corrections to $\frac{\Delta T_{\nu}}{T_{\nu}}$ vs
$T/m$ for $T>m$.

\end{document}